\documentclass{article}
\usepackage{ace,amsmath,graphicx,url,times, cite}
\usepackage{siunitx}
\usepackage[nolist]{acronym}
\usepackage{amssymb, amsmath, epsfig, rotating, multirow, url}
\usepackage{mathabx} 
\usepackage{epstopdf}
\usepackage{url}
\usepackage[draft]{hyperref}
\usepackage{xspace}
\usepackage{balance} 
\pdfoutput=1
\title{Reverberation time estimation on the ACE corpus using the SDD method}
\name{James Eaton and Patrick A. Naylor}
\address{Department of Electrical and Electronic Engineering, Imperial College London, UK}
\newcommand{\figsMidSent}{Figs.\@\xspace}
\newcommand{\etal}{\emph{et al.}\@\xspace}
\newcommand{\gHertz}[1]{\SI{#1}{\giga\hertz}}
\newcommand{\msec}[1]{\SI{#1}{\milli\second}}
\newcommand{\seco}[1]{\SI{#1}{\second}}
\newcommand{\dBel}[1]{\SI{#1}{\deci\bel}}
\newcommand{\sampleIndex}{n}
\newcommand{\sampleDep}{(\sampleIndex)}
\newcommand{\speech}{x}
\newcommand{\speechN}{\speech\sampleDep}
\newcommand{\noise}{\nu}
\newcommand{\noiseN}{\noise\sampleDep}
\newcommand{\noisyRevSpeech}{y}
\newcommand{\noisyRevSpeechN}{\noisyRevSpeech\sampleDep}
\newcommand{\convolvedWith}{\ast}
\newcommand{\RIR}{h}
\newcommand{\RIRN}{\RIR\sampleDep}
\newcommand{\SRMRID}{A}
\newcommand{\LollmannID}{B}
\newcommand{\WenID}{C}
\newcommand{\EatonID}{D}
\newcommand{\EatonRemID}{E}
\begin{document}
\begin{acronym}
\acro{ACE}{Acoustic Characterization of Environments\acroextra{. An IEEE challenge run by the SAP group at Imperial College}}
\acro{AI}{Articulation Index}
\acro{AIR}{Acoustic Impulse Response}
\acro{DR}{Douglas-Rachford}
\acro{DRR}{Direct-to-Reverberant Ratio}
\acro{NS}{Noise Suppression}
\acro{NSV}{Negative-Side Variance}
\acro{RIR}{Room Impulse Response}
\acro{RT}{Reverberation Time}
\acro{RTF}{Real-Time Factor}
\acro{SD}{Semantic Differential}
\acro{SDD}{Spectral Decay Distributions}
\acro{SNR}{Signal-to-Noise Ratio}
\acro{STFT}{Short Time Fourier Transform}
\acro{T60}[$T_\textrm{60}$]{Reverberation Time\acroextra{ to decay by $60$ dB}}
\acro{TI}{Texas Instruments, Inc.}
\acro{TIMIT}{\ac{TI}-\ac{MIT} speech corpus}

\acro{Dev}{Development}
\acro{Eval}{Evaluation}
\end{acronym}
\ninept
\maketitle
\begin{sloppy}
\begin{abstract}
\ac{T60} is an important measure for characterizing the properties of a room.
The author's \ac{T60} estimation algorithm was previously tested on simulated data where the noise is artificially added to the speech after convolution with a impulse responses simulated using the image method.
We test the algorithm on speech convolved with real recorded impulse responses and noise from the same rooms from the \ac{ACE} corpus and achieve results comparable results to those using simulated data.
\end{abstract}

\begin{keywords}
Reverberation time, speech processing, room impulse response
\end{keywords}
\acresetall
\section{Introduction}
\label{sec:intro}
The acoustic properties of a room can be characterized by its \ac{AIR}.
From a measured \ac{AIR}, the acoustic parameters of \ac{T60} and \ac{DRR} can be determined using methods such as~\cite{ISO_3382} and~\cite{Mosayyebpour2012}.
These parameters can help improve the performance of speech enhancement and speech recognition.
However, in practical situations such as mobile telecommunications, the \ac{AIR} is not usually available.
In such circumstances the parameters must be determined from the speech non-intrusively i. e. without prior knowledge of the room acoustics.
%
In addition, noise will always be present~\cite{Naylor2012} and must be taken into account.
Algorithms to determine acoustic parameters are typically developed and tested using simulated acoustic environments where an artificial \ac{RIR} is combined with noise either generated electronically, or recorded in a different acoustic environment.
The simulated acoustics typically use the following signal model:
\begin{equation}
\noisyRevSpeechN = \speechN \convolvedWith \RIRN + \noiseN, 
\label{eqn:noisySpeech}
\end{equation}
%
%
where $ \noisyRevSpeechN$ is the noisy reverberant speech, $\speechN$ is anechoic speech, $\RIRN$ is the \ac{AIR}, and $\noiseN$ is additive noise.

It can be seen from \eqref{eqn:noisySpeech} that any ambient noise that does not have the same \ac{AIR} as the speech will not produce a realistic simulation.
The \ac{ACE} Challenge provides a set of matched recorded \acp{AIR} and noises thus alleviating this problem, and gives an opportunity to evaluate existing algorithms on the new data.
The contribution of this paper is to evaluate the author's \ac{T60} estimation method in~\cite{Eaton2013} on the \ac{ACE} corpus alongside the methods of Wen {\etal}~\cite{Wen2008}, Falk {\etal}~\cite{Falk2010a}, and L\"{o}llmann {\etal}~\cite{Lollmann2010} as evaluated in Gaubitch {\etal}~\cite{Gaubitch2012}.

The remainder of the paper is organized as follows:
In Section~\ref{sec:review}, the author's \ac{T60} estimation method is reviewed.
In Section~\ref{sec:perfeval}, the performance evaluation is described.
In Section~\ref{sec:results}, the results of the experiments are discussed and in Section~\ref{sec:conc}, conclusions are drawn.

\section{Review of the Reverberation Time Estimation Method}
\label{sec:review}
The author's \ac{T60} reverberation time estimation method~\cite{Eaton2013} is based on the \ac{SDD} method~\cite{Wen2008}.
This method determines the gradient within frames of the log magnitude \ac{STFT} by time and frequency bins for a speech signal.
As matrix of gradients is obtained by time and frequency, one for each frame of time-frequency bins.
It was observed in~\cite{Wen2008} that variance of the negative gradients or decays at speech end-points within this matrix, known as the \ac{NSV} correlate well with the \ac{T60} of a speech signal.
It was subsequently observed in~\cite{Gaubitch2012} that the method exhibited a strong bias in noisy reverberant speech, and that the computational complexity was high.

The authors sought to improve on this method in two ways.
Firstly, by finding a method to reduce the bias in the presence of additive noise, and secondly to reduce the computational complexity.
This was achieved by reducing the number of \ac{STFT} frequency bins in a perceptually motivated approach by averaging across frequencies in Mel frequency bands thus reducing the number of terms in the variance computation.
This averaging process also reduces the susceptibility to noise because the noise is averaged out.
Further, the \ac{STFT} time-frequency bins used to perform the computation of the \ac{NSV} were selected based on an estimate of the \ac{SNR} of the noisy reverberant speech.
In addition, the gradients can be computed efficiently using the Moore-Penrose inverse~\cite{Moore1920}.
\section{Performance evaluation}
\label{sec:perfeval}
The performance evaluation was performed using the Evaluation stage software provided by the \ac{ACE} challenge.
The \ac{ACE} challenge corpus~\cite{Eaton2015a} comprises \num{4500} noisy reverberant speech files.
This is based on 5 male and 5 female talkers with 5 utterances each of different lengths of anechoic speech.
Three different \acp{SNR} are used: High (\dBel{18}), Medium (\dBel{12}), and Low (\dBel{-1}). 
Three different noise types are applied: Ambient which is the sound of the room with no speech, Fan, the sound of the room with one or more fans operating, and Babble, the sound of multiple talkers speaking simultaneously in the room reading from \acs*{TIMIT} passages or scientific papers.
\acp{RIR} from 5 different rooms each with two different microphone positions are convolved with the anechoic speech and then mixed with noise using the \emph{v\_addnoise} function~\cite{Brookes1997}.

Two versions of the algorithm in~\cite{Eaton2013} were tested.
The first version is the published version using the published training coefficients for the mapping of the \ac{NSV} to the \ac{T60}.
In the second version, the upper limit of training range of \acp{T60} for the simulated \ac{RIR} was increased from \msec{950} to \msec{1850} cover a larger range of \ac{T60} than the original implementation.
No training was performed on the \ac{ACE} corpus.
%

The method in~\cite{Wen2008} in the original implementation used QR factorization and did not exploit the opportunity to solve the least squares in a matrix covering all time frequency bins.
This resulted in a high computational cost when compared when compared with other methods.
The implementations of {\WenID}, {\EatonID} and {\EatonRemID} for the \ac{ACE} Challenge were all revised to use the Moore-Penrose inverse~\cite{Moore1920} and compute the gradients for all time-frequency bins in a matrix for each frequency band.
%

%
The authors' two methods are compared with the methods of~\cite{Falk2010a,Lollmann2010,Wen2008} which were the subject of the evaluation of Gaubitch {\etal}~\cite{Gaubitch2012}.
For brevity, the algorithms are henceforth lettered from {\SRMRID} to {\EatonRemID} as
Falk {\etal}~\cite{Falk2010a}, 
L\"{o}llmann {\etal}~\cite{Lollmann2010}, 
Wen {\etal}~\cite{Wen2008},
Eaton {\etal}~\cite{Eaton2013}, 
and
Eaton {\etal}~\cite{Eaton2013} trained on \acp{T60} up to \msec{1850}
respectively.
%

In addition to estimation performance, the \ac{RTF} of each method is compared.
Algorithms {\SRMRID} and {\LollmannID} were tested using Matlab on an Intel Xeon X5675 processor with a clock speed of \gHertz{3.07},
whilst algorithms {\WenID}, {\EatonID}, and {\EatonRemID} were tested using Matlab on an Intel Xeon E5-2643 processor with a clock speed of \gHertz{3.30}.  
These two processors have similar levels of performance.
\section{Results}
\label{sec:results}
Results in the three noise types, Ambient, Fan, and Babble are shown in {\figsMidSent}~\ref{fig:ACE_Ambient},~\ref{fig:ACE_Fan}, and~\ref{fig:ACE_Babble} respectively.
\begin{figure}[!ht]
	\centerline{\epsfig{figure=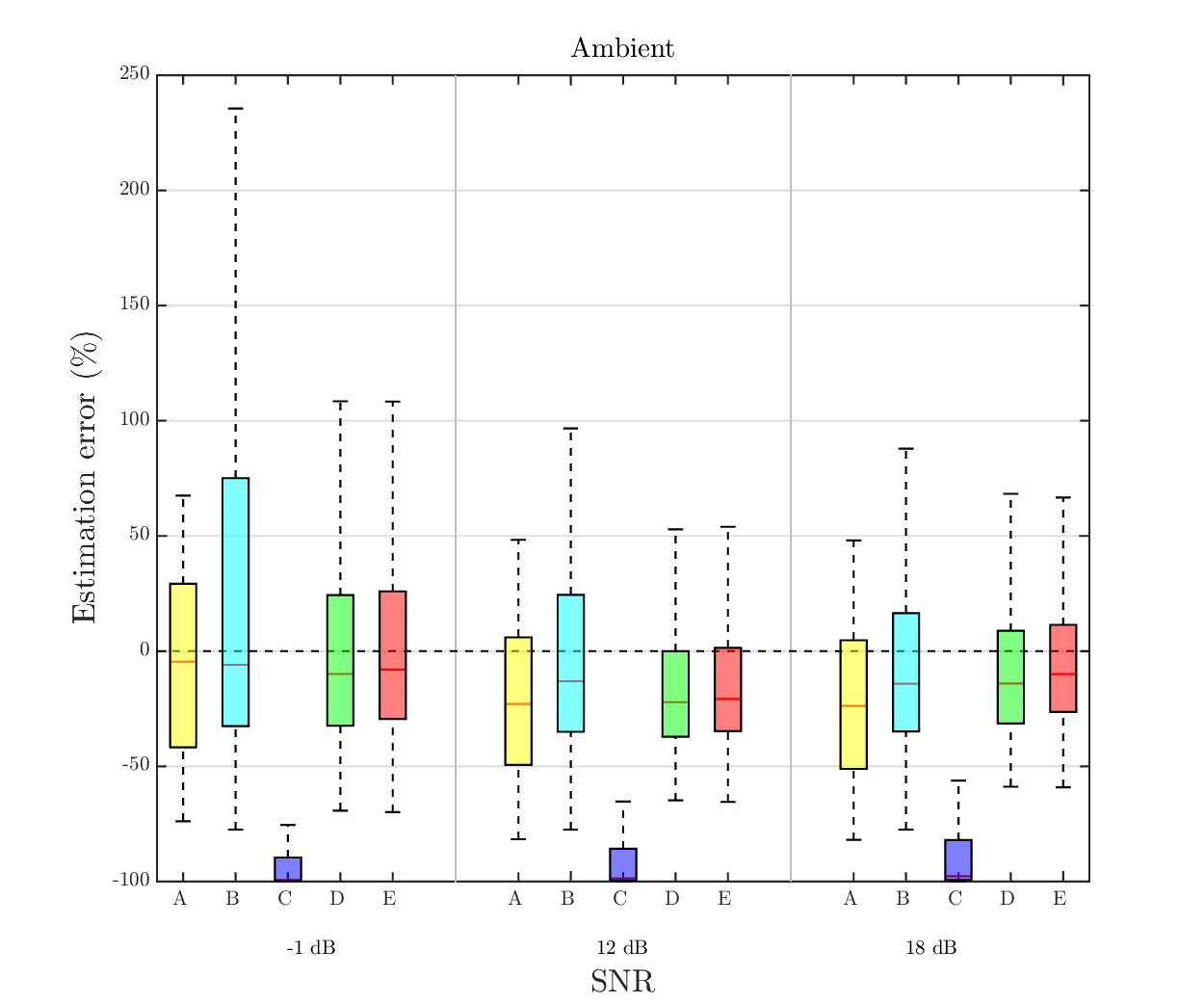,
	width=80mm,viewport=30 0 552 468,clip}}%
	\caption{{\ac{T60} estimation error for Ambient noise for computation methods {\SRMRID} to {\EatonRemID} for low (\dBel{-1}), medium (\dBel{12}) and high (\dBel{18}) \acp{SNR}}}%
\label{fig:ACE_Ambient}%
\end{figure}%
\begin{figure}[!ht]
	\centerline{\epsfig{figure=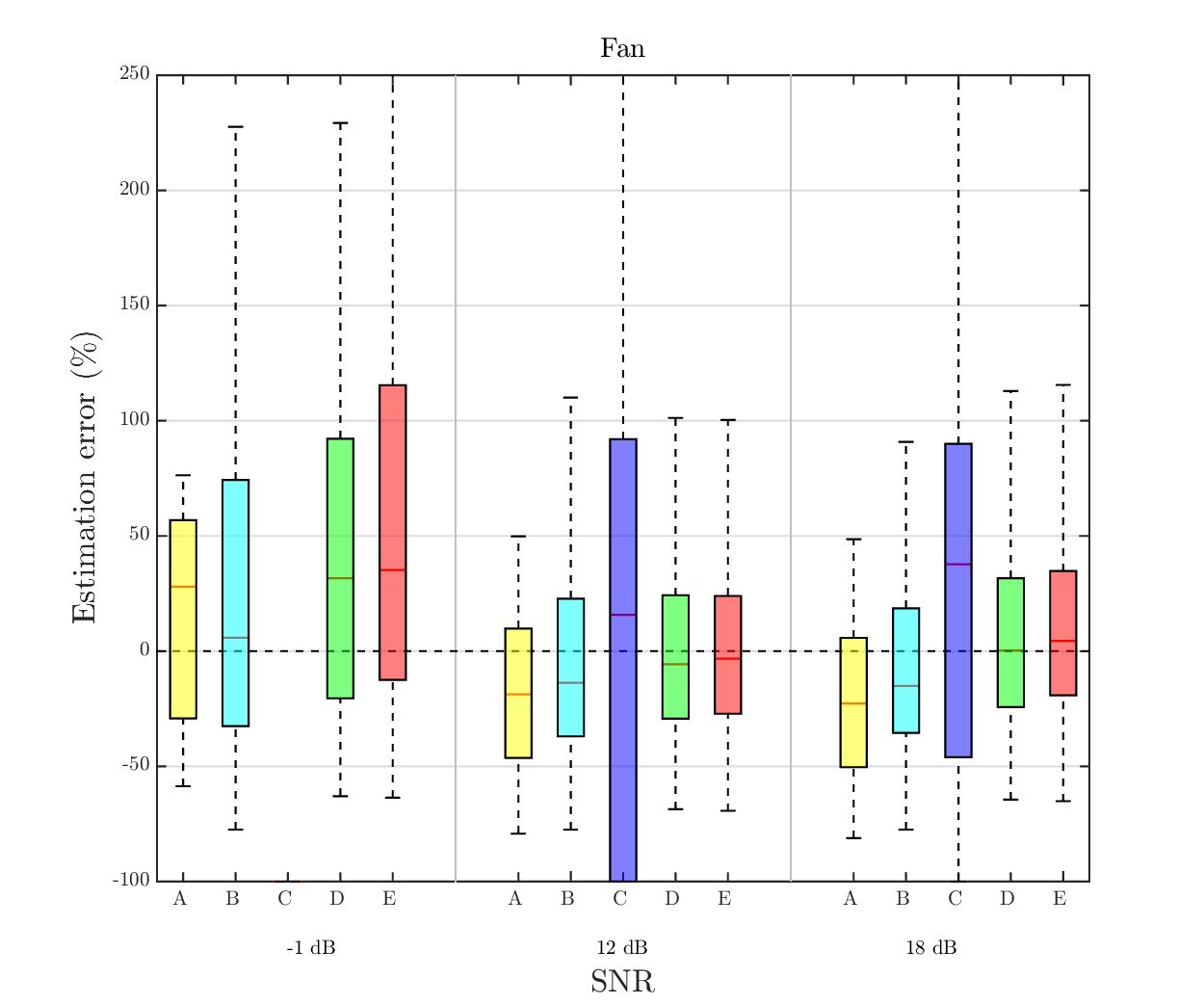,
	width=80mm,viewport=30 0 552 468,clip}}%
	\caption{{\ac{T60} estimation error for Fan noise for computation methods {\SRMRID} to {\EatonRemID} for low (\dBel{-1}), medium (\dBel{12}) and high (\dBel{18}) \acp{SNR}}}%
\label{fig:ACE_Fan}%
\end{figure}%
\begin{figure}[!ht]
	\centerline{\epsfig{figure=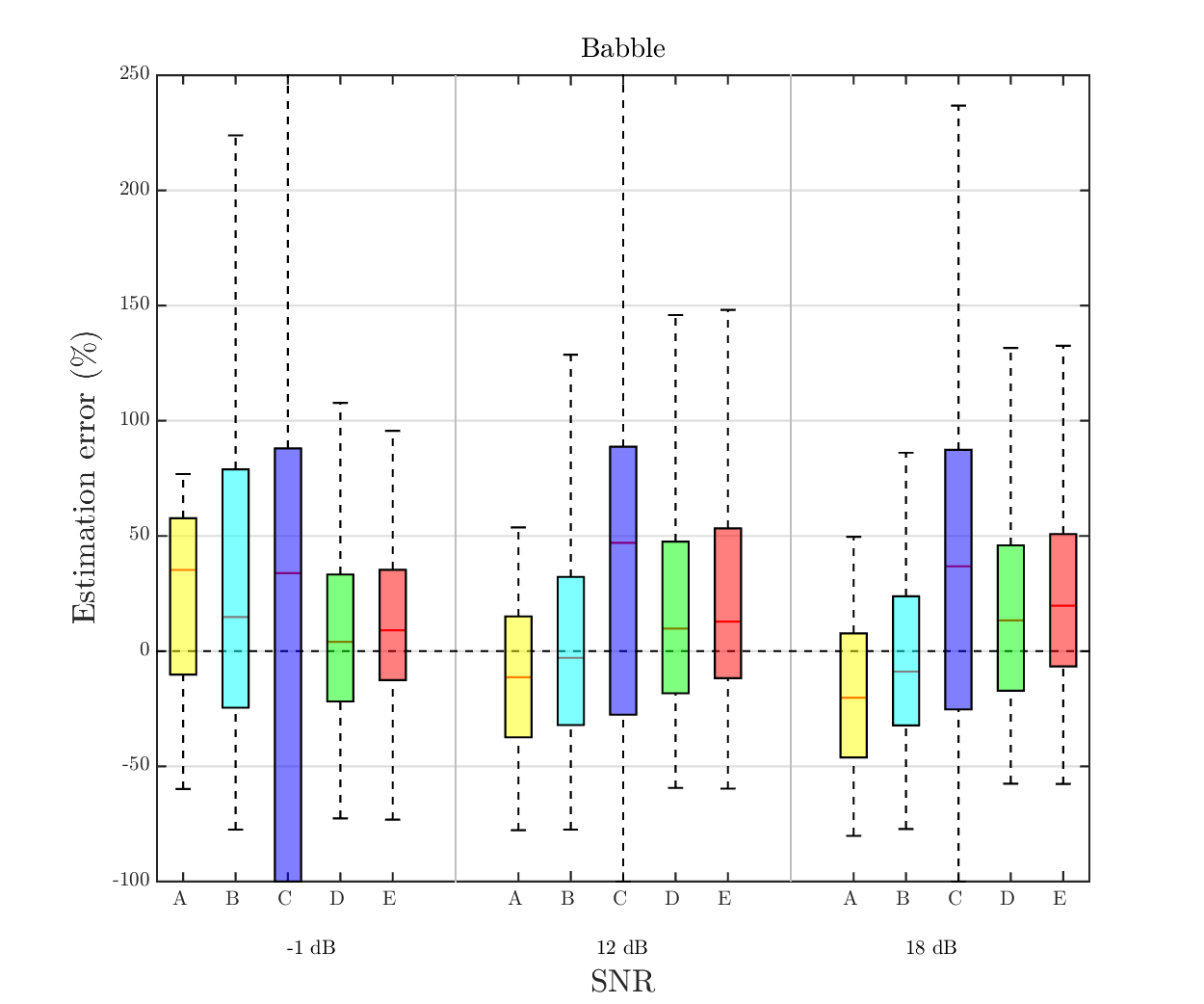,
	width=80mm,viewport=30 0 552 468,clip}}%
	\caption{{\ac{T60} estimation error for Babble noise for computation methods {\SRMRID} to {\EatonRemID} for low (\dBel{-1}), medium (\dBel{12}) and high (\dBel{18}) \acp{SNR}}}%
\label{fig:ACE_Babble}%
\end{figure}%
On each box, the central mark is the median, the edges of the box are the \num{25}th and \num{75}th percentiles,
the whiskers extend to the most extreme data points not considered outliers.
Outliers are not shown.

The results show that method {\WenID} does not cope well in noise confirming the results of~\cite{Gaubitch2012}.
This is because as the noise level increases, it causes the average gradient in each frame of \ac{STFT} time-frequency bins to approach zero resulting in very small \ac{NSV}.
This is the same effect as having a very long \ac{T60}.
The algorithm will in general therefore tend to overestimate, however, the relationship between the \ac{NSV} and the \ac{T60} is based on a trained mapping function which can result in a negative value of \ac{T60}, and in such circumstances the algorithm returns an estimate of \seco{0}.

In babble noise, the end-point decays of the babble will have the same \ac{T60} as the speech.
Under these circumstances method {\WenID} gives a better estimate.

The remaining methods give similar levels of performance with fan noise being the most challenging environment in which to estimate at low \acp{SNR} for all algorithms.

%

Table~\ref{tab:performance} shows the \ac{RTF} for each algorithm computed by dividing the total CPU time used to process all \num{4500} noisy reverberant speech files in the \ac{ACE} Evaluation dataset by the combined length of all the noisy reverberant speech files.
\begin{table} [htb] 
\caption{\label{tab:performance} {Comparison of \ac{RTF}}} 
\vspace{2mm} 
\centerline{ 
\begin{tabular}{|c|c|c|c|c|} 
	\hline 
{\SRMRID}~\cite{Falk2010a}   &{\LollmannID}~\cite{Lollmann2010}   &{\WenID}~\cite{Wen2008} &{\EatonID}~\cite{Eaton2013} &{\EatonRemID}~\cite{Eaton2013} \\ 
	\hline	\hline 
\num{0.457}	&\num{0.131}    &\num{0.0219}  &\num{0.0166}       &\num{0.0153}     \\ 
	\hline 
\end{tabular}}
\end{table}
This results in a much lower \ac{RTF} in the Matlab implementation.
The perceptually motivated averaging using Mel-spaced frequency bands in {\EatonID} and {\EatonRemID} gives a further reduction in \ac{RTF}.
%
%
\section{Conclusion}
\label{sec:conc}
The \ac{ACE} Challenge has provided an opportunity to evaluate the author's algorithm on real recordings of \acp{AIR} and noise in contrast to the previously reported performance which was on simulated data.
Also, the speech in the \ac{ACE} corpus is free-running and less uniform than the \acs{TIMIT} database used in~\cite{Eaton2013}, and so is representative of conversational speech rather than read speech.
The estimation performance compares well with existing methods, and has a low \ac{RTF} making it suitable for real-time applications.
\balance
\bibliographystyle{IEEEtran}
\bibliography{../SapBibTex/sapref}
\label{sec:bibliography}
\end{sloppy}
\end{document}